\documentclass[11pt]{article}

\usepackage{moriond}
\usepackage[dvips]{graphicx}
\usepackage{epsfig}

\def\Journal#1#2#3#4{{#1} {\bf #2}, #3 (#4)}

\def\PLB{{\em Phys. Lett.}  B}
\def\PRL{\em Phys. Rev. Lett.}

\begin{document}
\vspace*{4cm}
\title{SEARCHES FOR NEW PHYSICS IN PHOTON AND JET FINAL STATES}

\author{ M. Jaffr\'e\\on behalf of the CDF and D0 collaborations}

\address{LAL, Universit\'e Paris-Sud, CNRS/IN2P3,\\
91898 Orsay Cedex, France}

\maketitle\abstracts{
Recent results from searches of physics beyond the standard model in $p\bar{p}$
collisions are reported, in particular, reactions involving high transverse
momentum photons or jets in their final state.
Data analyzed by the CDF and D0 experiments at the Run II of the Tevatron correspond
to integrated luminosities between $1$ and $2 fb^{-1}$ depending of the analyses.
}

\def\MET{{\mbox{$E\kern-0.57em\raise0.19ex\hbox{/}_{T}~$}}}
\def\METnoSpace{{\mbox{$E\kern-0.57em\raise0.19ex\hbox{/}_{T}$}}}

\newcommand{\sq}{\mbox{$\tilde{q}$}}
\newcommand{\sg}{\mbox{$\tilde{g}$}}
\newcommand{\st}{\mbox{$\tilde{t}$}}
\newcommand{\xo}{\mbox{$\tilde{{\chi}}^0_1$}}
\newcommand{\xpm}{\mbox{$\tilde{{\chi}}^{\pm}_1$}}

\section{Introduction}

At an energy frontier collider, the usual way to search for indices of
physics beyond the standard model (SM) 
is to look for the collisions with the highest momentum-transfer particles.
Typically, one chooses a particular model, and the event selection is
optimized to enhance its contribution against the SM expectation.
The absence of any deviation in data provides  a limit on the production cross-section times the branching
ratio for the channel under study, which is then translated into exclusion limits
in the parameter space of this model.
However, by nature, a new phenomena is unknown, and it exists a lot of models at disposal.
This is the reason which motivates the ''signature-based'' search strategy which casts a wider
look for deviations to the SM.

Both strategies will be reported here for final states with photons and jets.

\section{Randall-Sundrum (RS) graviton}
Many models with extra spatial dimensions have been proposed to solve the
 hierarchy problem. In the RS model~\cite{RS}, the SM brane and the Planck brane are
separated by an extra dimension with a warped geometry.
Only the graviton is allowed to propagate in this extra dimension.
It appears as Kaluza-Klein (KK) towers in the SM brane.
This model has only 2 parameters:
$M_1$, the mass of the lowest KK excited mode, and
$k/M_{Pl}$, a dimensionless coupling constant whose value should lie between 0.01 and 0.1.

\begin{figure}[htb]

  \begin{minipage}[t]{0.33\linewidth}
    \includegraphics[width=5.0cm]{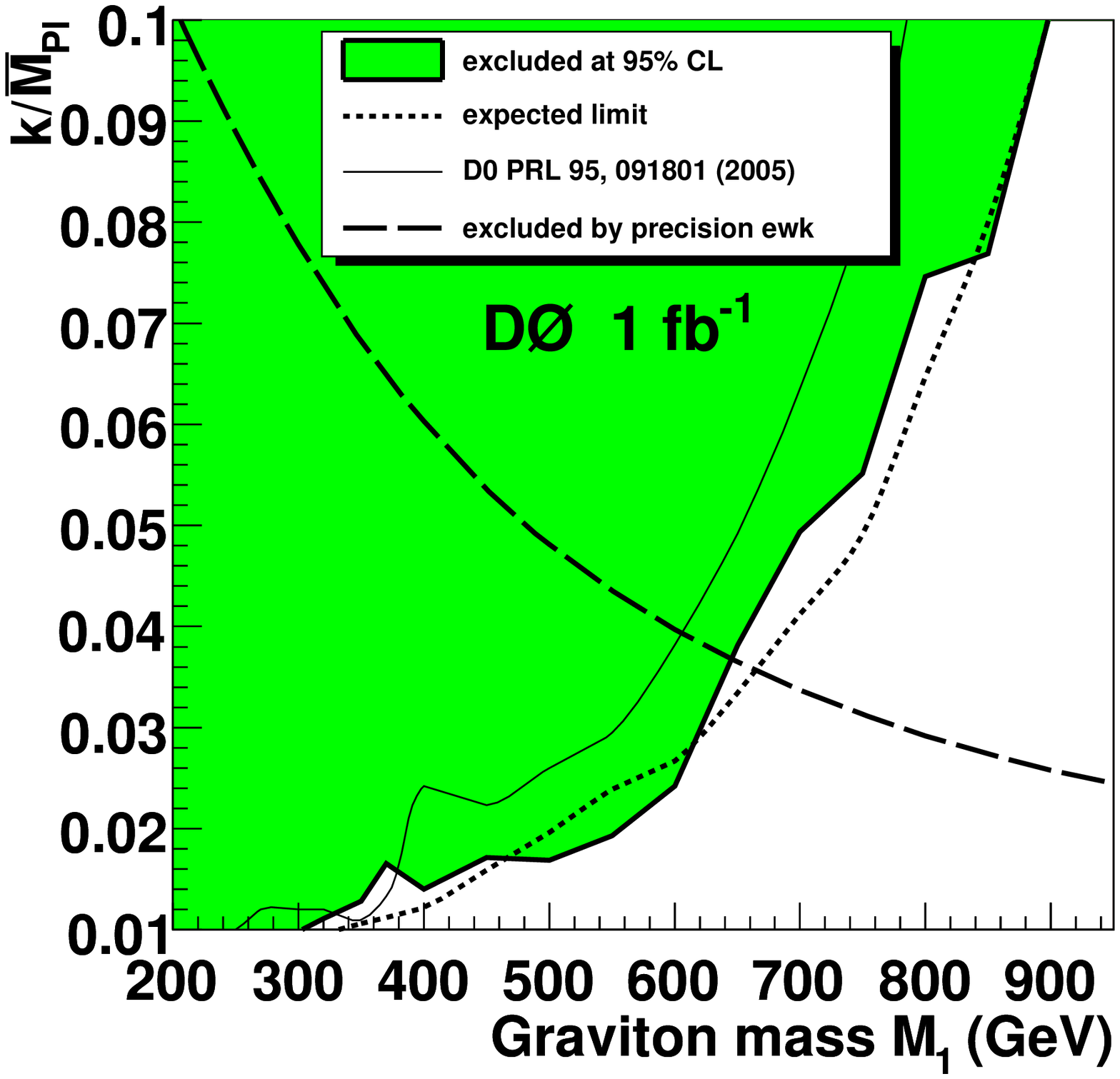}
    \caption{\label{fig:RSresult}
   95$\%$ C. L. upper limit on $k/\overline{M}_{Pl}$ versus graviton mass compared with the RS expected limit.}
  \end{minipage}
  \hspace{5mm}
  \begin{minipage}[t]{0.66\linewidth}
  \begin{picture}(110,50)
  \put(-2,1.5){\hbox{
    \includegraphics[width=5.5cm]{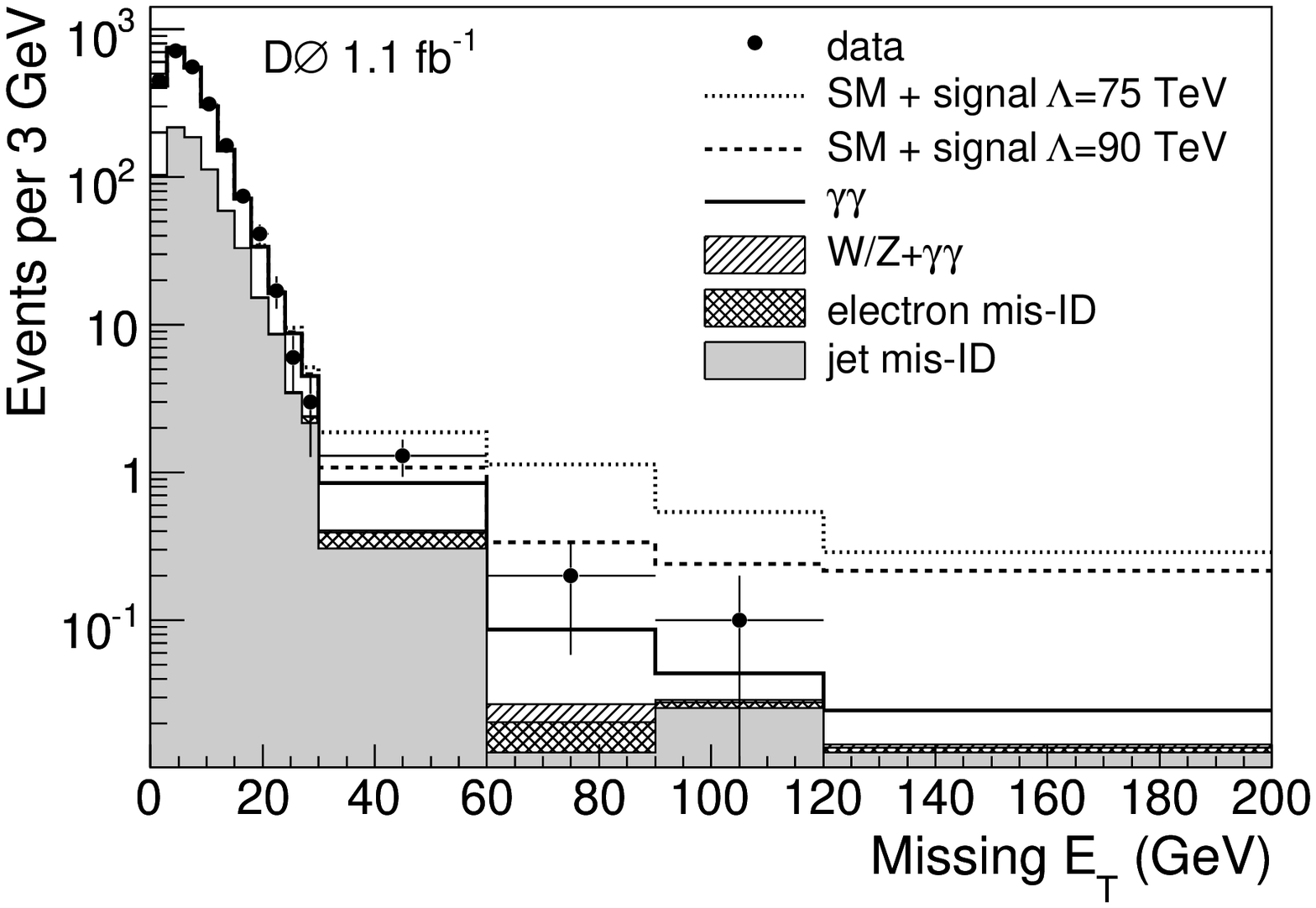}
    \includegraphics[width=5.0cm]{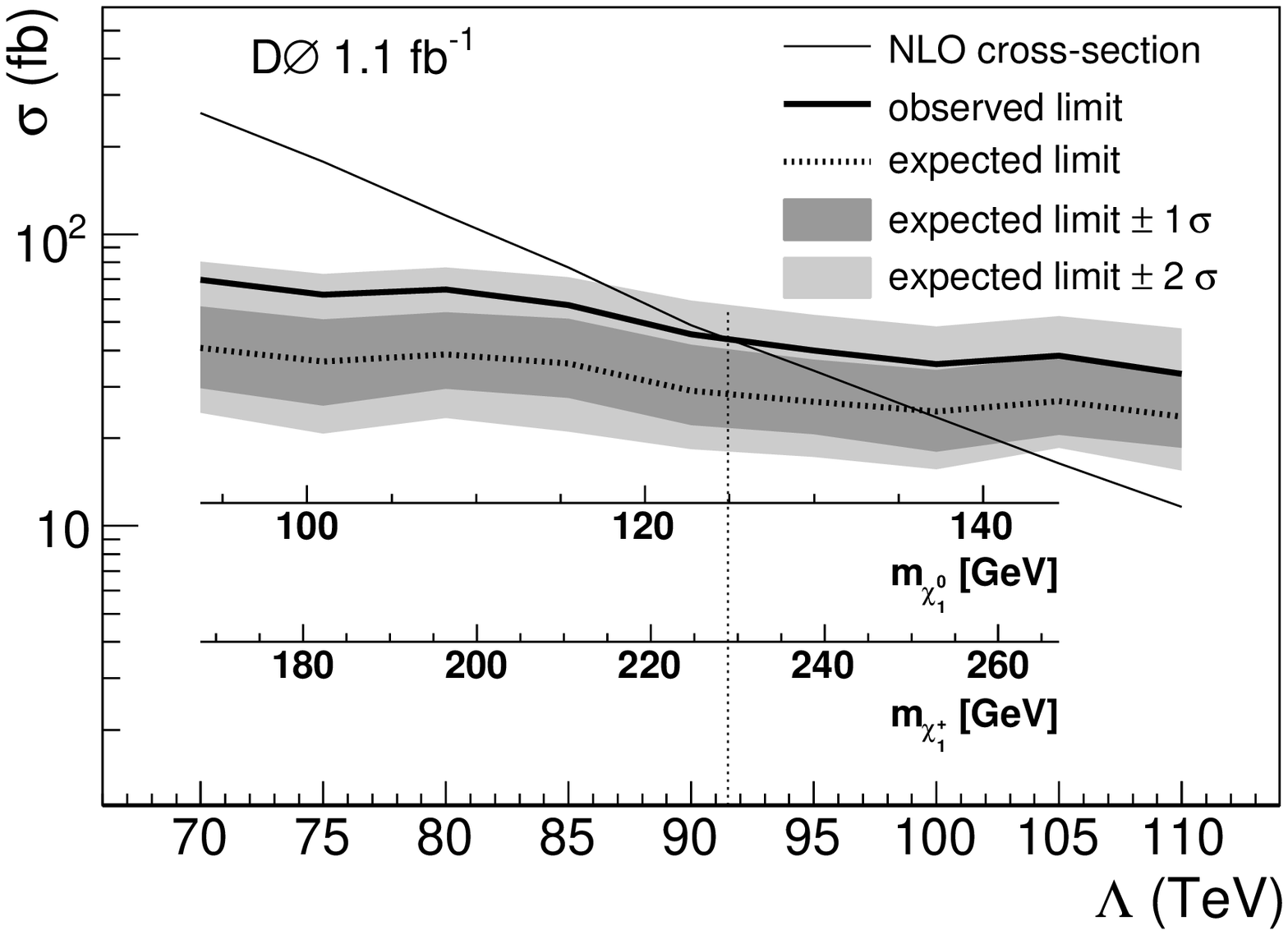}
    }}
  \end{picture}
    \caption{\label{fig:GMSBresult}
   The \MET distribution in $\gamma\gamma$ data with the various background
contributions (left).
 Predicted cross section for the Snowmass Slope model versus $\Lambda$. The observed and expected 95\% C.L. limits are also shown (right).}
  \end{minipage}

\end{figure}

KK towers couple to any boson or fermion pairs.
CDF~\cite{cdf-RS} looks separately at $\gamma\gamma$ and $ee$ final states, whereas D0~\cite{d0-RS}
looks at both final states at once as they look similar in the electromagnetic
calorimeter.
Because of the spin 2 of the graviton, the ratio of the branching ratios
to $\gamma\gamma$ and $ee$ final states  is 2.
Both experiments have analyzed about the same amount of data ($\sim 1~fb^{-1}$),
and found no excess of events over the SM predictions ( Drell-Yan and QCD where
 jets are misidentified as photons) excluding graviton masses below $900~GeV/c^2$
for $k/M_{Pl}=0.1$.
Fig.~\ref{fig:RSresult} shows the excluded contour in the 2D parameter space as measured by D0.

\section{Gauge mediated SUSY breaking(GMSB)}
SUSY~\cite{susy} is a broken symmetry. Experimental signatures are determined through
the manner and scale of the SUSY breaking.
In the GMSB scenario, the lightest supersymmetric particle (LSP) is the gravitino,
a very light and weakly interacting particle. The next to lightest supersymmetric
particle (NLSP) is 
assumed in this analysis to be 
the neutralino which decays into the LSP and a photon.
Assuming R-parity conservation~\cite{rpar}, SUSY particles are pair produced and
the experimental signature  will be 2~photons and missing energy from the 2 gravitinos.
To get a quantitative result, the ''Snowmass Slope SPS 8'' model~\cite{snowmass} is considered.
All the GMSB parameters\,\footnote{The messenger mass $M_m=2\Lambda$, 
the number of messengers $N_5=1$, $tan(\beta)=15$, $\mu>0$.}
are fixed as a function of the effective energy scale $\Lambda$ of SUSY breaking.

In this event topology, the SM background is the $Z\gamma\gamma$ production where the
$Z$ boson decays into neutrinos.
There is also important instrumental background from events with real $\MET$ ($W$ boson production)
and fake $\MET$ ( QCD where jets are misidentified as photons).
Fig.~\ref{fig:GMSBresult} (left) shows the $\MET$ distribution.
The observed distribution agrees well with the SM prediction; the entire 
spectrum is then used to set limits on the GMSB production cross section.
Fig.~\ref{fig:GMSBresult} (right) shows the $95\%$ C.L. cross section limit as a 
function of the effective scale $\Lambda$ obtained by D0~\cite{d0-GMSB}.
The observed limit on the signal cross section is below the prediction of
 the Snow-mass Slope model for $\Lambda < 91.5~TeV$, or for gaugino masses $m_{\xo}<125~GeV/c^2$ and
 $m_{\xpm} < 229~GeV/c^2$.   

\section{Large Extra Dimensions}
The hierarchy problem can also be solved by postulating the existence of $n$ new large extra dimensions
as proposed first by Arkani-Hamed, Dimopoulos and Dvali~\cite{add} (ADD); the extra volume serves to
dilute gravity so that it appears weak in our 3D world as
the graviton is the only particle allowed to propagate in the extra space.
If the extra dimensions are compactified in a torus of radius R, according to the Gauss law, one can relate
the fundamental Planck mass scale $M_D$, R, the Planck mass and the number of extra dimensions by the relation 
$M_{Planck}^{2}= 8 \pi M_D^{n+2} R^{n}$, allowing $M_D$ to be compatible with the electroweak scale.

In this model, the graviton can be  produced directly in the reaction $q \bar{q} \rightarrow G \gamma$; G will remain
undetected leaving a signature with a single photon and $\MET$.

The only SM background is the $ Z \gamma$ production where the Z boson decays into a neutrino pair.
In addition to the usual instrumental background coming from misidentification of electrons or jets into photons,
the event topology is rather sensitive to a contribution from beam halos and cosmics where muons produced photons by bremsstrahlung.

To fight the latter background, both experiments had to develop specific tools in addition to the usual ones based on the EM shower
profile.
Special hit finders in the tracker starting from the EM cluster increase the track veto efficiency.
In addition, D0 uses a EM pointing tool thanks to its preshower detector, and CDF the timing system built within its EM calorimeter.

\begin{figure}[htb]
  \begin{picture}(160,60)
  \put(0,2){\hbox{
 \includegraphics[width=7.5cm]{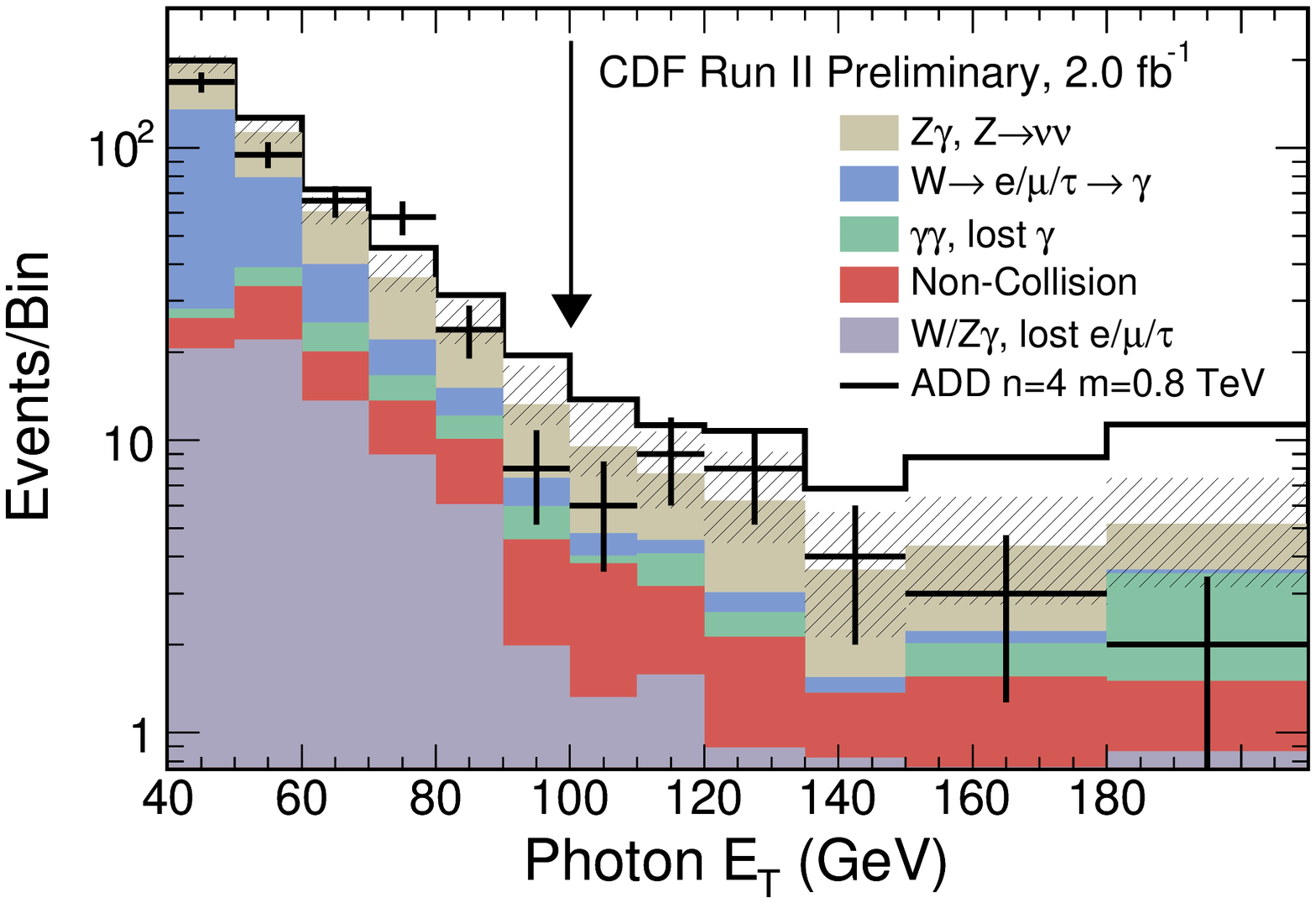}
 \includegraphics[width=7.0cm]{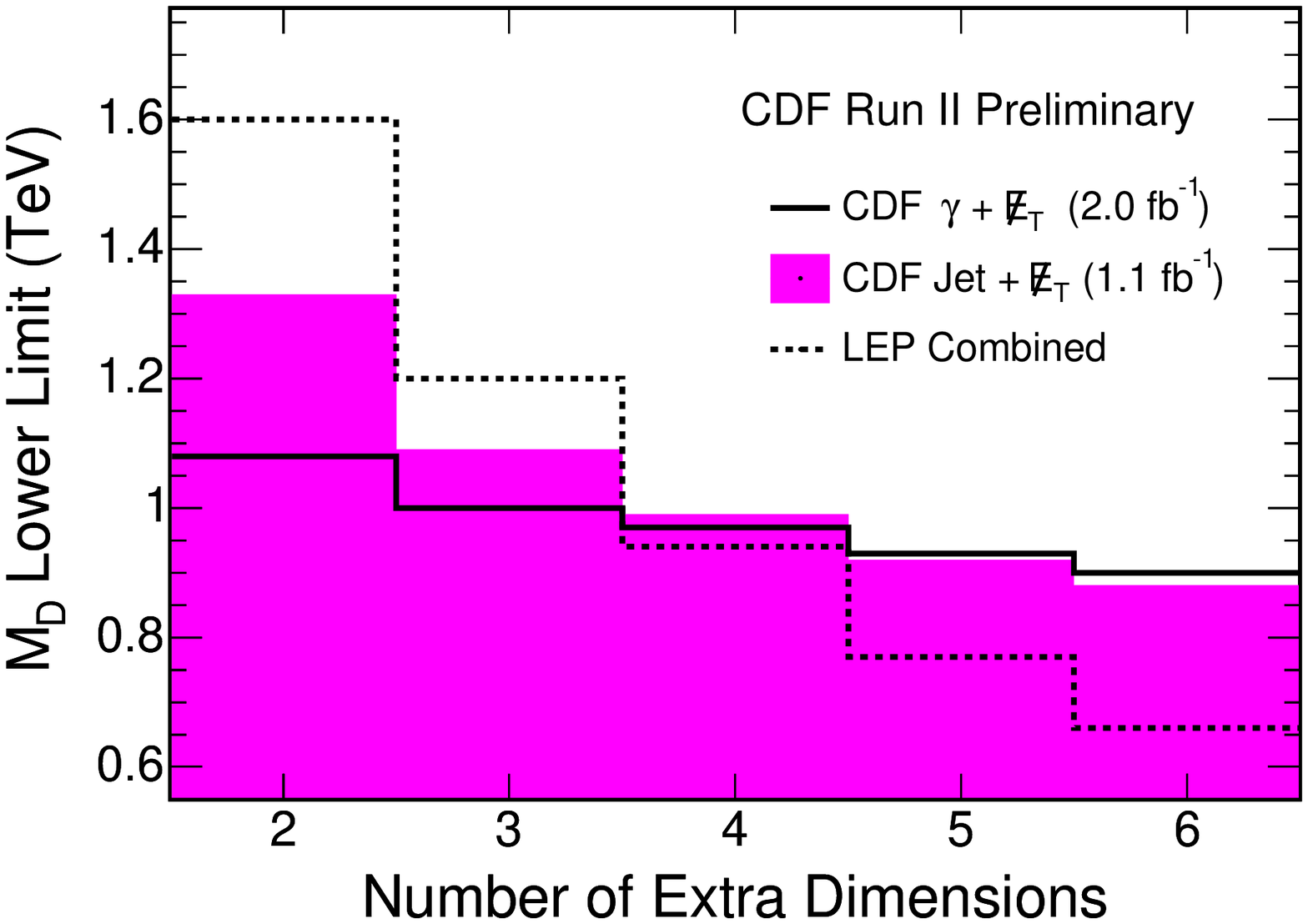}
    }}
  \end{picture}
  \caption{\label{fig:LEDresult}
   The left figure shows the \MET distribution in the CDF monophoton search.
The signal expected from the ADD model (n=4, m=0.8TeV) is added on top of the SM backgrounds.
 The right figure shows the exclusion limits for the ADD model obtained in this analysis, in comparison with the CDF jet and \MET result 
and LEP combined result.}
\end{figure}

The results of CDF which has analysed about $2 fb^{-1}$ of data, twice as much  as D0~\cite{d0-monoPhoton}, are displayed in Fig.~\ref{fig:LEDresult}.
The left plot shows a good agreement for the photon transverse between data and the sum of the various backgrounds.
This allows to set limits on the fundamental scale $M_D$ (right plot) as a function of the number of extra dimensions.
For $n>4$ the limits are comparable with the limits obtained in 
the monojet search, and better than the LEP combined result~\cite{lep-led}. 

\section{Squarks and gluinos, stops}
\subsection{Squarks and gluinos}
\begin{figure}[htb]
  \begin{picture}(160,75)
  \put(0,5){\hbox{
 \includegraphics[width=7.5cm]{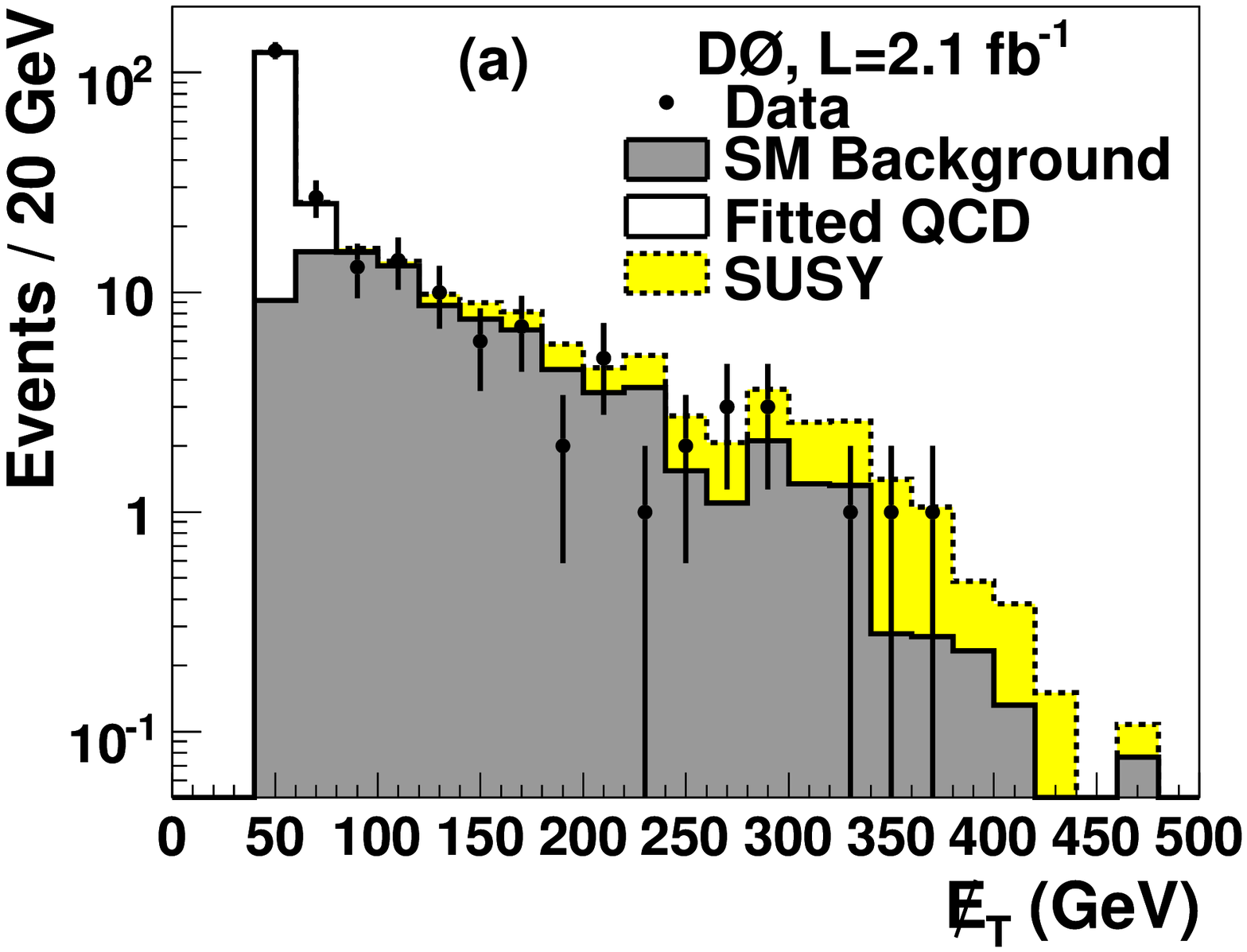}
 \includegraphics[width=8.25cm]{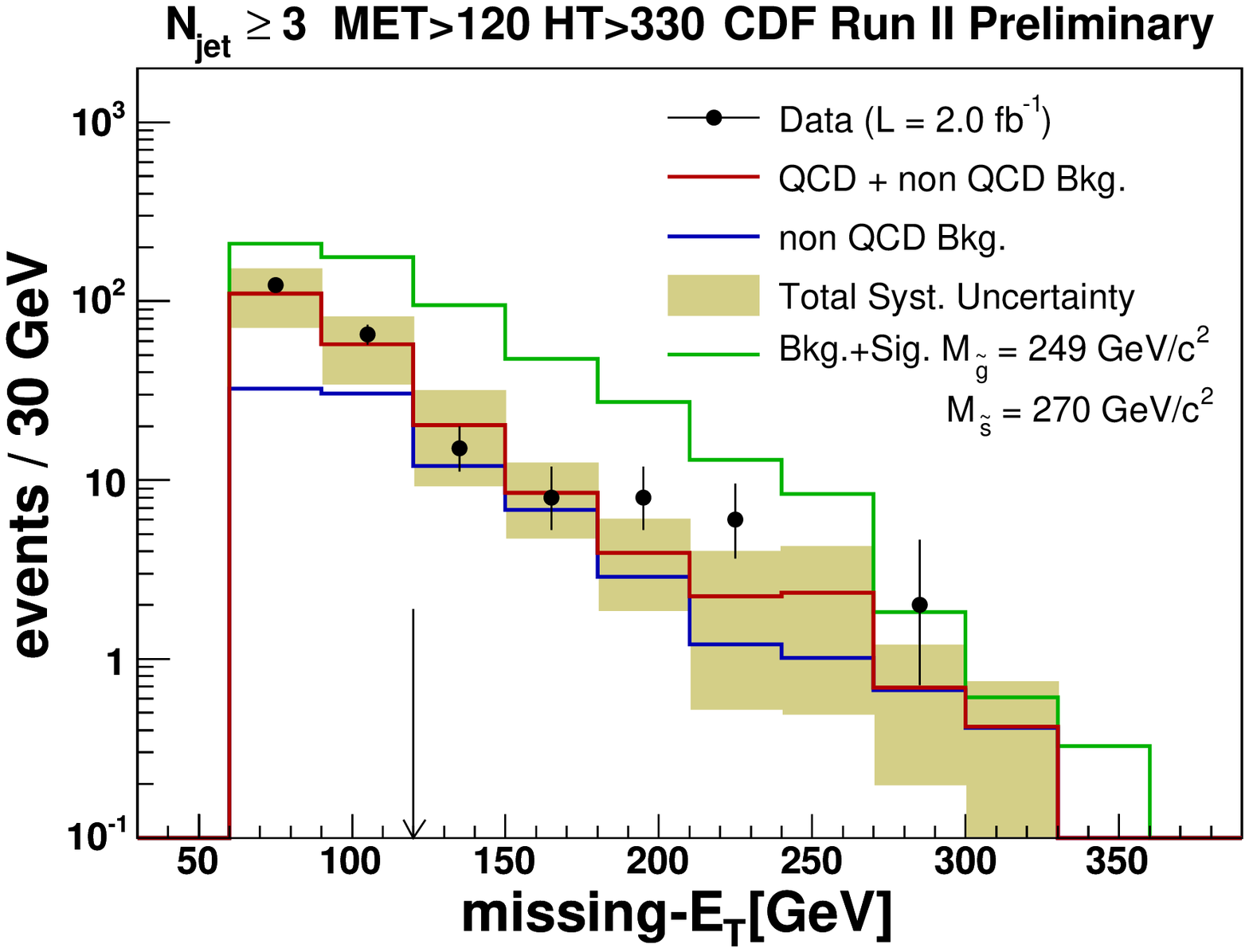}
    }}
  \end{picture}
  \caption{\label{fig:sqglmET}
   Distributions of $\MET$ after applying all analysis criteria except the one on $\MET$ for the 
   ``dijet'' (D0 left) and ``3-jets'' (CDF right) squark-gluino analyses; 
   data (points with error bars) and 
   the cumulated contributions from SM background, QCD background  and signal MC are shown.}
\end{figure}

Squarks and gluinos can be copiously produced at the Tevatron if they are
sufficiently light.
The analysis is performed within the mSUGRA model~\cite{msugra}.
The final state is composed of jets with a large $\MET$ due to the two escaping 
neutralinos, assumed to be the LSP.
According to the relative mass of squarks and gluinos different event topologies are to be expected.
If squarks are lighter than gluinos,  a ''dijet'' topology is favored.
On the contrary if  squarks are heavier than gluinos, the final state
contains at least 4 jets.
Finally, the jet multiplicity is at least $3$ if squarks and gluinos have 
similar masses.
After a common event preselection, the three topologies have been studied and
optimised separately.
The left plot on Fig.~\ref{fig:sqglmET} shows the D0 $\MET$ 
distribution obtained in the ''dijet'' search, the right one is obtained
by CDF in the ''3-jet'' search.
D0 has analyzed $2.1~fb^{-1}$ of data without finding any excess over
the SM predictions.
It allows to extend the exclusion domain in the squark gluino plane
(Fig.~\ref{fig:SQGLresult}).
Using the most conservative hypothesis, D0~\cite{d0-sqgl}(CDF~\cite{cdf-sqgl})
excludes a  gluino lighter than $308(290)~GeV/c^2$.
    
\subsection{Stop}

\begin{figure}[htb]
  \begin{minipage}[t]{0.45\linewidth}
    \includegraphics[width=7.cm]{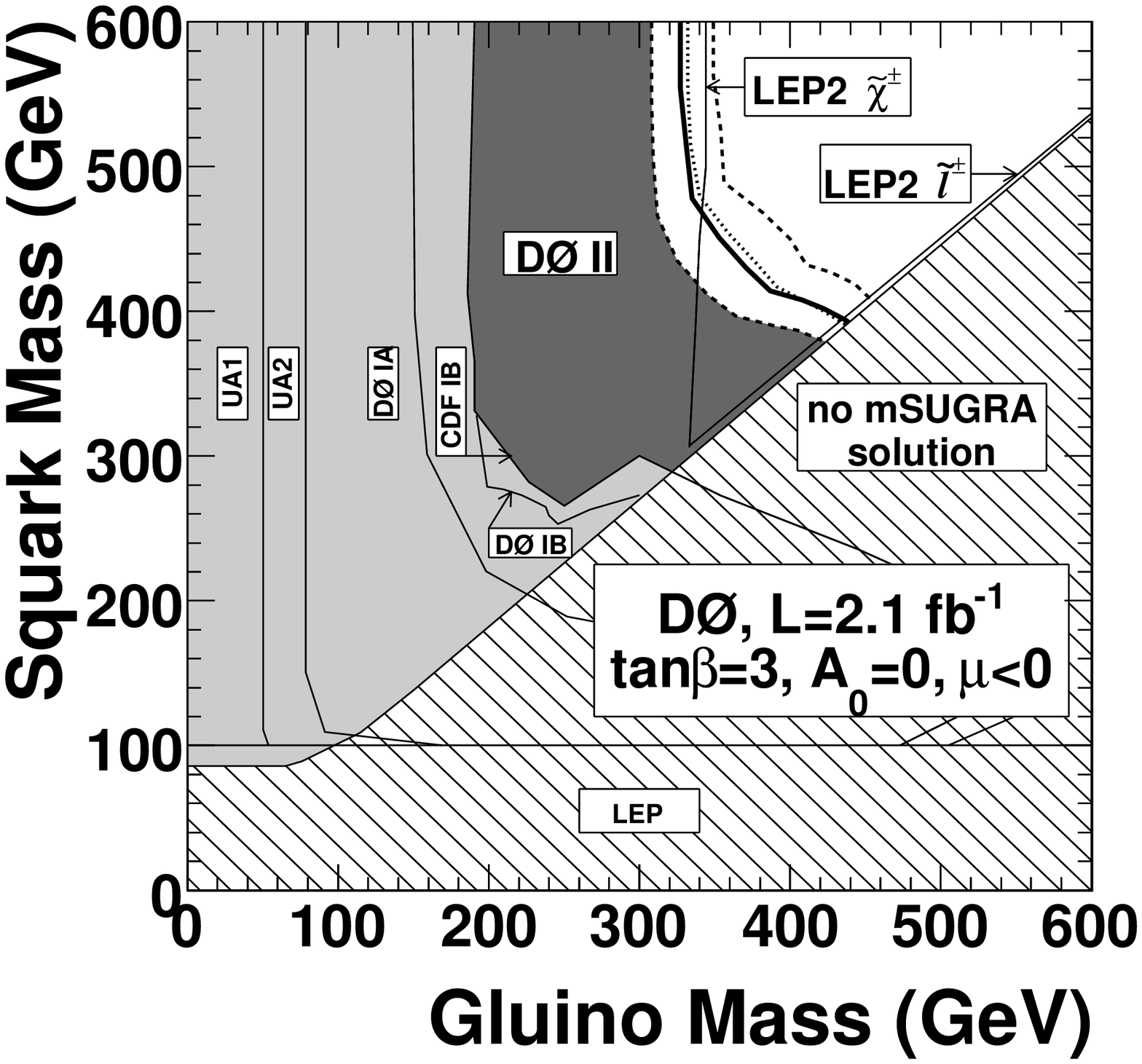}
    \caption{\label{fig:SQGLresult}
    Excluded region in the squark and gluino mass plane; newly excluded domain
    by D0 is shown in dark shading.
    The region where no mSUGRA solution can be found is shown hatched.}
  \end{minipage}
  \hspace{5mm}
  \begin{minipage}[t]{0.45\linewidth}
    \includegraphics[width=7.cm]{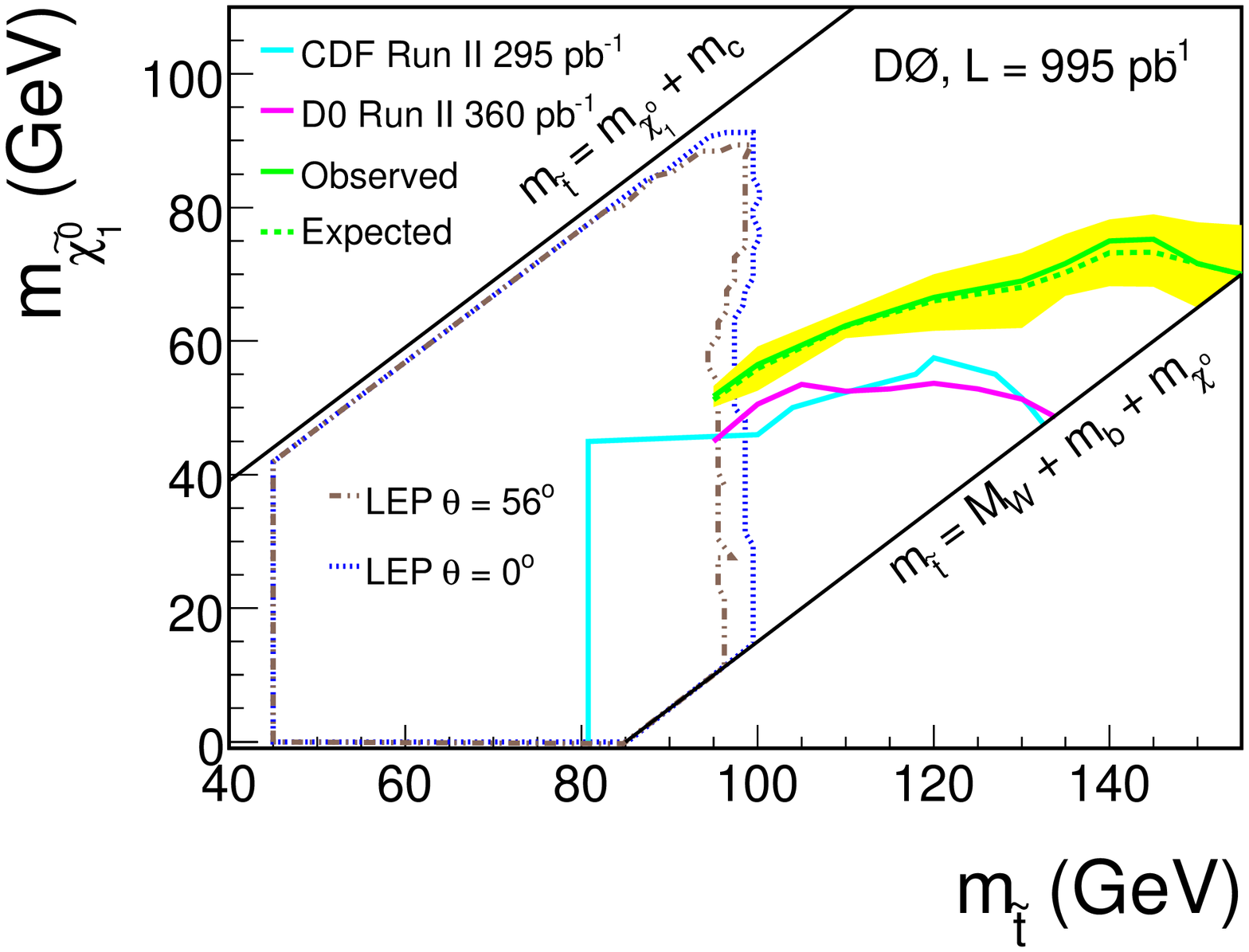}
    \caption{\label{fig:STOPresult}
    Region in the {\st}--{\xo} mass
plane excluded at the $95\%$ C.L. by the D0 search. The yellow band represents
 the theoretical uncertainty on the scalar top quark pair production cross
 section due to PDF and renormalization and factorization scale choice.}
  \end{minipage}
\end{figure}

Due to the large Yukawa coupling, there could be a large mixing in the 3rd generation
of squarks.
The lighest of the 2 stops could be the lightest squark and even the NLSP.
Furthermore if its mass is less than the sum of the masses of the b quark, 
the W boson, and the neutralino, the dominant decay mode is $\st \rightarrow c\xo$,  
a flavor changing loop decay, assumed to be $100\%$ in the analysis.
The final state will then be $2$ acoplanar charm jets and $\MET$.
The analysis proceeds with $2$ jets detected in the central part of the
detector with a loose heavy quark tag for one of them.
No excess of events has been observed~\cite{d0-stop} in about $1~fb^{-1}$ of data,
which provides a lower limit for the stop mass at $149~GeV/c^2$ for a
neutralino mass of $63~GeV/c^2$ (Fig.~\ref{fig:STOPresult}).  

\section{Signature-based searches}
\subsection{Search for anomalous production of di photon events}
\begin{figure}[htb]
  \begin{minipage}[t]{0.45\linewidth}
    \includegraphics[width=7.cm]{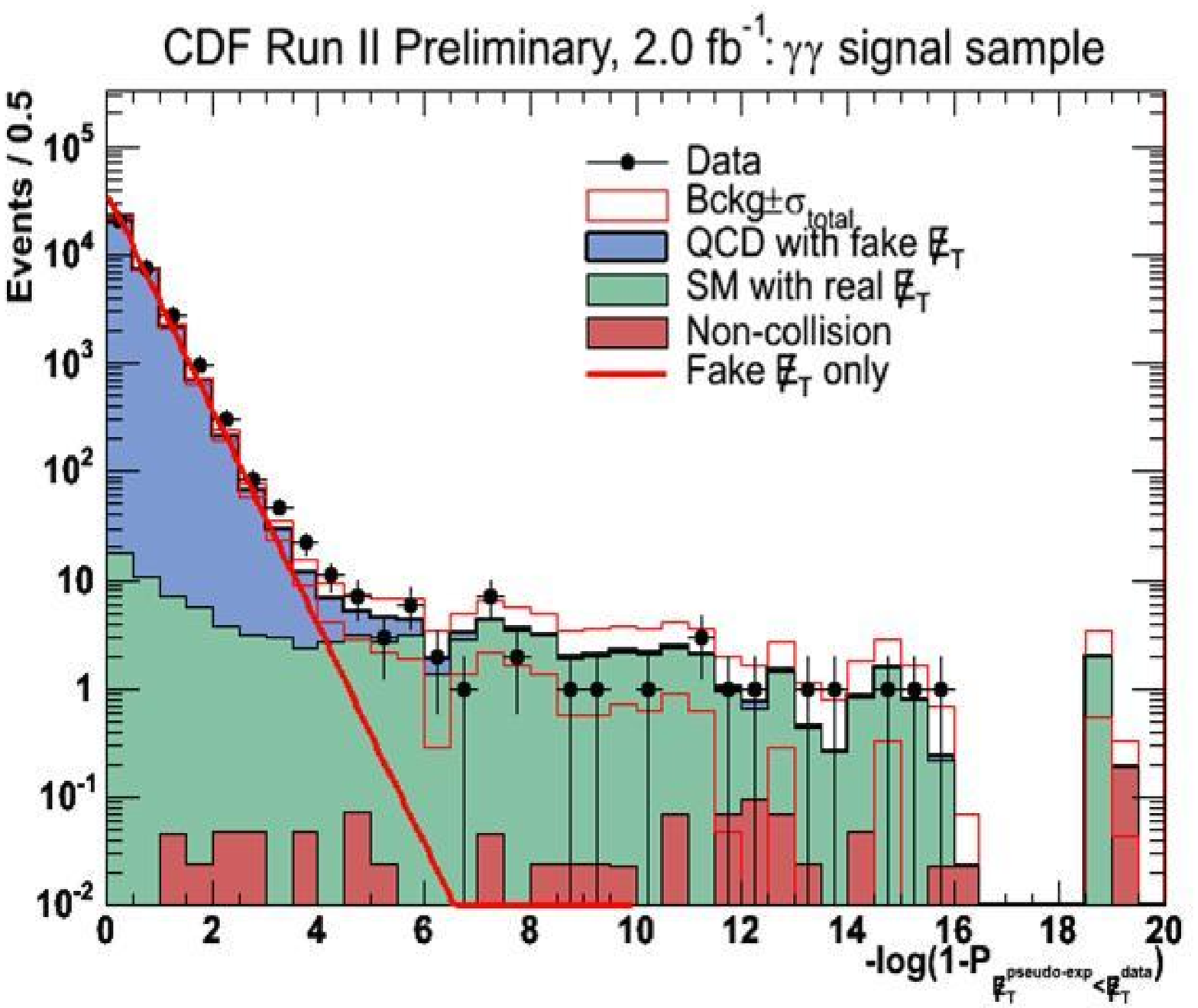}
    \caption{\label{fig:cdfdiphoton}
     Distribution of missing transverse energy significance for diphoton candidates.}
  \end{minipage}
  \hspace{5mm}
  \begin{minipage}[t]{0.45\linewidth}
    \includegraphics[width=7.cm]{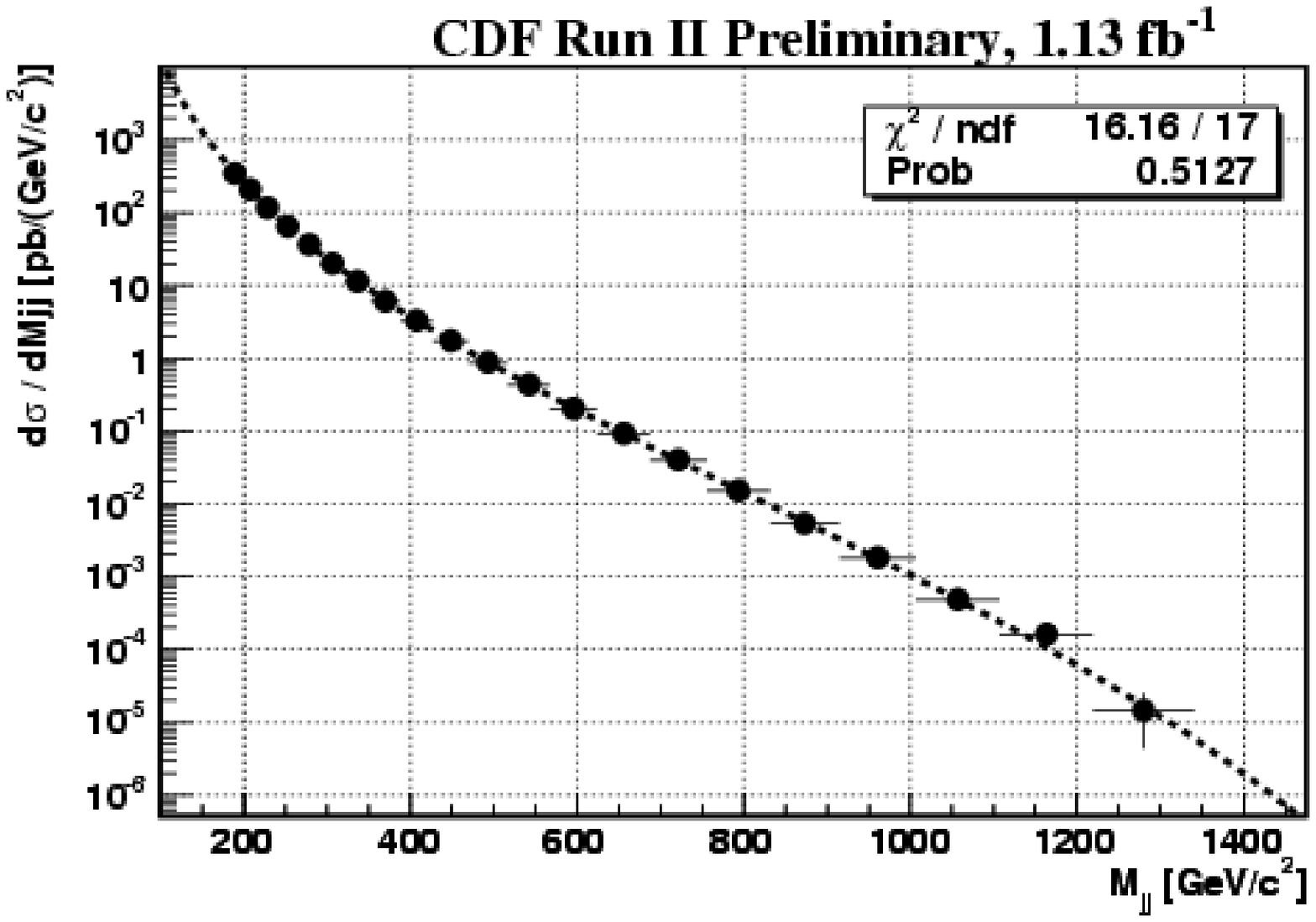}
    \caption{\label{fig:dijetmass}
    The measured dijetmass spectrum and results of the fit to the parametrization form~\ref{eq:dijet-param}.}
  \end{minipage}
\end{figure}

In its quest for ``signature-based'' excess, CDF has searched for anomalous
 production of events in the $\gamma\gamma+ \MET$ topology.
In this analysis, use is made of the $\MET$ resolution model.
The aim of this model is to discriminate events with large mismeasured $\MET$
from events with real $\MET$.
It has been shown to provide a better background rejection power than a simple
$\MET$ cut.
This model is based on the assumption that individual particle's energy resolution
has Gaussian shape proportional to particle's $\sqrt{E_T}$. Only two sources 
of fake $\MET$ are considered : soft unclustered energy ( from underlying 
event and multiple interactions ), and jets. The latter is responsible for
most of the $\MET$ as it is collimated energy in contrast to the former which
is spread out all over the calorimeter.
According to this model, each event is given a $\MET$ significance value.
Most of the QCD background is eliminated by requiring a significance above 5,
leaving  only the expected number of SM events with real $\MET$(Fig.~\ref{fig:cdfdiphoton}), and not much room for an extra signal.

\subsection{Search for dijet mass resonances}
 Many classes of models beyond the SM predict the existence of new massive
particles decaying into 2 partons which would appear as  resonances in
the dijet mass spectrum. Such classes include excited quarks, techniparticles,
new W' or Z' bosons, RS graviton,...
 Jets are reconstructed by the cone-based midpoint jet algorithm~\cite{run2cone}
 with a cone radius of $0.7$ and have central rapidity ($|y|<1$).
CDF has analysed about $1.1~fb^{-1}$ of data and measured 
the dijet differential cross section (Fig.~\ref{fig:dijetmass}).
The spectrum is fitted by the smooth parametrization :
\begin{equation}
\frac{d\sigma}{dm} = p_0 (1-x)^{p_1} /x^{p_2+p_3\log(x)}, \hspace{1cm} x=m/\sqrt(s).
\label{eq:dijet-param}
\end{equation}
This parametrization is found to fit well the dijet spectra from PYTHIA and HERWIG MC events as well as from NLO pQCD.
As no evidence for existence of a new massive particle is observed, limits
on new particle production cross sections can be derived as a function
of the dijet mass. These limits are then translated into mass exclusion
limits, see Table~\ref{tab:dijet-exclusions}. 

\begin{table}[htb]
    \caption{\label{tab:dijet-exclusions}
    Mass exclusion ranges for several models.}
    \begin{center}
    \tiny
    \begin{tabular}[b]{|l|c|}
    \hline
    &\\
    Model description & Observed mass exclusion range \\
      & ($GeV/c^2$)\\
    \hline
    Excited quark ($f=f'=f_s=1$) & 260-870\\[2mm]
    Color octet technirho & \\
    (top-color-assisted-technicolor couplings) & 260-1110\\[2mm]
    Axigluon and flavor universal coloron &\\
    (mixing of 2 SU(3)'s cot(theta)=1) & 260-1250\\[2mm]
    E6 diquark & 290-630\\[2mm]
    W' (SM couplings) & 280-840\\[2mm]
    Z' (SM couplings) & 320-740\\[2mm]
    \hline
    \end{tabular}
   \end{center}
\end{table}

\section{Conclusions}
No hints of physics beyond the SM have been found so far. 
As the Tevatron is continuing to provide experiments with more data to analyze,
the quest for indices will be pursued by CDF and D0.
Some analyses presented in this talk have already been published, for the others,
further details can be found at:
\begin{description}
\item[CDF] http://www-cdf.fnal.gov/physics/exotic/exotic.html
\item[D0] http://www-d0.fnal.gov/Run2Physics/WWW/results/np.htm
\end{description}

\section*{Acknowledgments}
The author would like to thank the CDF and D0 working groups for providing
the material for this talk, and the organizers of the {\em Rencontres} for a
very enjoyable conference and the excellent organization.

\end{document}